# The use of registry data to extrapolate overall survival results from randomised controlled trials

**Reynaldo Martina[1,5], Keith Abrams[1], Sylwia Bujkiewicz[1], David Jenkins[1,6], Pascale Dequen[1,7], Michael Lees[2], Frank A. Corvino[3], and Jessica Davies[4]**

[1] Department of Health Sciences, University of Leicester, University Road, Leicester, LE1 7RH, UK

[2] WWHEOR (Markets), Bristol Myers Squibb, 3 rue Joseph Monier, 92500 Rueil-Malmaison, France

[3] Genesis Research LLC, 5 Marine View Plaza, Hoboken, New Jersey, USA 07030

[4] Personalized Health Care Data Science-Product Development, F. Hoffmann-La Roche Ltd, 6 Falcon Way, Shire Park, Welwyn Garden City, AL7 1TW, UK

[5] Department of Biostatistics, University of Liverpool, 1-5 Brownlow Street, Liverpool, L69 3GL, UK

[6] School of Health Sciences, University of Manchester, Oxford Road, Manchester, M13 9PL, UK

[7] Evidence Synthesis/Health Economics, Visible Analytics Ltd., Union Way, Witney, OX28 6HD, UK

on behalf of GetReal Workpackage 1

## Abstract

**Background:** Pre-marketing authorisation estimates of survival are generally restricted to those observed directly in randomised controlled trials (RCTs). However, for regulatory and Health Technology Assessment (HTA) decision-making a longer time horizon is often required than is studied in RCTs. Therefore, extrapolation is required to estimate long-term treatment effect. Registry data can provide evidence to support extrapolation of treatment effects from RCTs, which are considered the main sources of evidence of effect for new drug applications. A number of methods are available to extrapolate survival data, such as Exponential, Weibull, Gompertz, log-logistic or log-normal parametric models. The different methods have varying functional forms and can result in different survival estimates.

**Methods:** The aim of this paper was to use registry data to supplement the relatively short term RCT data to obtain long term estimates of effect. No formal hypotheses were tested. We explore the above parametric regression models as well as a nonparametric regression model based on local linear (parametric) regression. We also explore a Bayesian model constrained to the long-term estimate of survival reported in literature, a Bayesian power prior approach on the variability observed from published literature, and a Bayesian Model Averaging (BMA) approach. The methods were applied to extrapolate overall survival of a RCT in metastatic melanoma.

**Results:** The results showed that the BMA approach was able to fit the RCT data well, with the lowest variability of the area under the curve up to 72 months with or without the SEER-Medicare registry.

**Conclusion:** the BMA approach is a viable approach to extrapolate overall survival in the absence of long-term data.

# 1 Introduction

Pre-marketing authorisation estimates of survival are generally restricted to those observed directly in randomised controlled trials (RCTs). For health technology assessment (HTA) it is important to assess the relative effectiveness of products for a longer period of time than is usually assessed in conventional RCTs. Extrapolation is a common approach in survival analysis to assess the probability of long-term survival following an intervention. Registry data can provide evidence to support extrapolation of treatment effects from RCTs, the main sources of evidence of effect for new drug applications, because registries generally contain patient cohorts with longer follow-up time than is feasible to obtain from RCTs and therefore they could contribute to obtaining better long term survival estimates. However, there is no clear recommendation on how to use RCT data in conjunction with different types of registry data (individual patient data (IPD) or summary data) for extrapolation purposes. There is also uncertainty regarding the best model to use for extrapolation.

## 1.1 Previous work and limitations

Various parametric models have been studied to extrapolate overall survival (OS) from RCTs. One of the most commonly used methods to extrapolate the results of RCTs is fitting a parametric survival model to the available RCT data, and use the parametric survival model to extrapolate beyond the follow-up time of the RCT (Latimer, 2013). The most commonly used parametric survival models include the exponential, Weibull, Gompertz, log-logistic and log-Normal regression models (Jackson *et al.* 2016).

Latimer (2013) evaluated survival analysis methods commonly used for economic evaluations, in particular those used for extrapolation when IPD are available. In the review by Latimer (2013), the survival analyses used for the technology appraisals conducted by the National Institute for Health and Care Excellence (NICE) were evaluated to demonstrate the limitations of survival analyses submitted for HTAs. The limitations include the lack of justification of the chosen survival models and the limited range of models considered. The issue with the models used for extrapolation is that different parametric models have different properties and as such, may yield different estimates of long-term survival. To aid the selection of the best model for systematic and transparent extrapolation of survival data, a criterion-based approach was provided by Tremblay *et al.* (2015). Five criteria were provided to evaluate a number of techniques used to extrapolate OS and progression free survival (PFS):

(i) Test the proportional hazard assumption, followed by evaluating criterion (ii) to (v) to select the best model

(ii) Fit extrapolated hazard function, including visual inspection of the extrapolation. Prioritise models where the hazard patterns seem reasonable

(iii) Select parametric model with lowest Akaike Information Criterion (AIC) and Bayesian Information Criterion (BIC) to demonstrate goodness-of-fit in the period before extrapolation

(iv) Account for uncertainty when selecting the best model since high uncertainty would be indicative of poor robustness

(v) Compare the pre and post extrapolation Area Under the Curve (AUC). A rule-of-thumb was used: check that the incremental average survival per month does not exceed the value pre-extrapolation.

In the example case study in metastatic breast cancer, discussed in Tremblay *et al.* (2015), the Accelerated Failure Time parametric model assuming a Gamma distribution and including a treatment covariate was the best of the models used for extrapolation of OS. For PFS, the Kaplan-Meier (KM) approach performed better. Limitations of the proposed criterion for selection include the step wise approach itself, to first test the proportional hazards assumption. Expert input regarding the underlying assumptions from past studies/data may be sufficient to guide which models to use since the statistical test may provide conflicting results of what is known from the data. It could therefore be argued that these steps need not be followed in chronical order. Expert knowledge based on past experience could lead to the *a priori* exclusion of some classes of models. On the other hand, viable models may be ignored because the results of the statistical test indicated a false negative result for those models.

In a recent publication by Guyot *et al*. (2016) on the extrapolation of survival curves from cancer trials, the Surveillance, Epidemiology and End Results (SEER) database (control arm only) was combined with other external information to extrapolate survival of RCTs using a single model for all data. The standard parametric models were applied with and without the use of the external data. A flexible spline model was also applied because the parametric models were unable to fit both the RCT data and the SEER data. Limitations of the proposed method include the subjectivity of the opinion regarding the long-term effect and the technical difficulty in using spline models when the available data are sparse. However, this is not specific to the proposed method. The authors confirm that the use of external data could contribute to obtaining "better" (or more appropriate) survival estimates.

### 1.2    Aim

We aimed to combine IPD from a RCT (Robert *et al*. 2011) with IPD from real world data and with summary data from published real world cohort studies (Joosse *et al*. 2011) to extrapolate the results from the RCT. We followed the systematic approach as outlined in the five criteria advocated by Tremblay *et al*. (2015). The ideas are similar to those from Guyot (2016), but we included IPD as well as summary data from registries both with and without the corresponding regression models being constrained by the external data. Different methods were considered to extrapolate overall survival (OS) from the RCT. The commonly used parametric models were applied to extrapolate the OS results as well as a nonparametric model based on local linear regression (Li and Racine, 2004). We also conducted a Bayesian power prior analysis. Chen and Ibrahim (2006) showed the advantage of the power prior approach to estimate the power parameter (variance) especially when multiple historical datasets are available. Ibrahim *et al*. (2015) described the advantages of using power prior approach over other informative prior distributions in general linear models, survival models and random effect models. Finally, we also considered a Bayesian Model Averaging (BMA) approach. An advantage of the BMA approach is that it accounts for both parameter and model uncertainty (Hoeting *et al*., 1999). The SEER database linked with Medicare claims data was used to obtain relevant treatment information needed for extrapolation of the RCT data.

The manuscript is constructed as follows: in the methods section the example data are described as well as the models used for extrapolation. This section also describes the different techniques, in particular, the so-called non-informative extrapolation, extrapolation using combined sources of evidence and extrapolation using registry data. A constraint regression analysis and a BMA approach are also described. The results for each approach are described in similar order in the results section as they appear in the methods section. The methods were applied to extrapolate the results of an RCT in metastatic melanoma. In the discussion section further discussion is provided on the use of real world data (RWD) for extrapolation of overall survival. It should be explicitly noted that extrapolation was applied on the overall survival results of the individual treatments separately. Also, due to the availability of registry data for standard of care (SOC) only, the constrained regression and Bayesian power model approach were applied on the SOC arm only.

## 2    Methods

The commonly used parametric models for extrapolation of RCT results are applied to extrapolate the OS results. To provide an alternative to parametric extrapolation, a nonparametric alternative based on local linear regression (Li and Racine, 2004) is also used.
A Bayesian power prior approach and Bayesian Model Averaging approach are also presented as alternatives. The following section provides a more detailed description of the data available and the methods applied.

### 2.1 Available data used to illustrate methods

*Randomised controlled trial data*

IPD were obtained from an RCT in untreated patients with metastatic melanoma (Robert *et al.*, 2011). Patients received SOC or an experimental treatment. A total of 502 patients were randomised in a 1:1 active to placebo ratio. Patients were followed up for 48 Months. Figure 1 illustrates the survival probabilities for each of the treatment arms included in the trial.

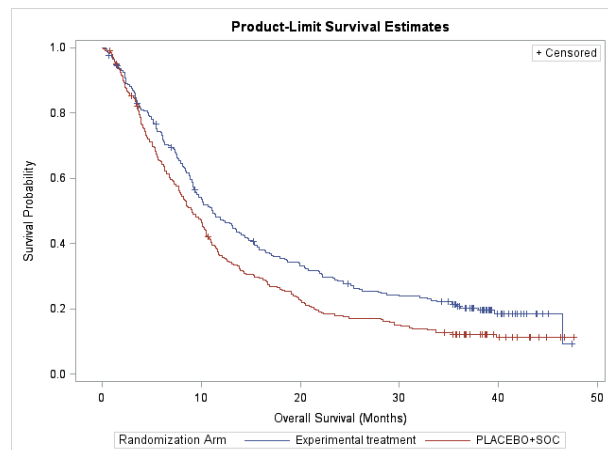

**Figure 1** Survival probability for the experimental treatment and standard of care

*Data from literature*

A literature search was performed to identify studies that included patients with metastatic melanoma similar to those included in the RCT. Two studies were identified that included results useful for purpose of the extrapolations:

Joosse *et al* (2011) published long-term estimates of survival of patients from the Munich cancer registry. A total of 11,774 patients were selected from this registry, which includes patients who were diagnosed with cutaneous melanoma. The study reported 1321 events (11.9%) of lymph node metastasis and 1602 events (13.7%) of distant metastasis. Total follow-up time was 120 months and median follow-up time was 80 months.

The published long-term survival estimates from Joosse *et al.* (2011) were used to assess the plausibility of the extrapolation strategies through visual inspection of the extrapolated results and are presented in Figures 3a, 3b, 5, 6 and 7 for reference purposes only. The lower of the two tick marks, in the figures mentioned above, at 80 months represents the survival probability of lymph node metastasis. The higher tick mark at 80 months (displayed at 81 months for clarity) represents the survival probability for distant metastasis.

In Altomonte *et al.* (2013) data were collected from metastatic melanoma (stage III/IV) patients who failed or did not tolerate previous treatment for metastatic melanoma in Italy and were enrolled in an expanded access program for the experimental treatment in the RCT. The published results included KM curves of patients followed up for 48 months. The KM curves were used to reconstruct IPD using the approach proposed by Guyot *et al.* (2012). As a result, these IPD could be combined with IPD data from the RCT to extrapolate long-term survival.

*SEER-Medicare cohort*

The SEER registry linked with Medicare claims (SEER-Medicare database) was used to obtain relevant treatment information needed for extrapolation of the RCT data. The SEER-Medicare database is the linkage of two large US population-based data sources that provide patient-level information on Medicare beneficiaries with cancer. A real world cohort of elderly patients (>66 years of age) with treated primary metastatic melanoma, reflecting the target patient population of the trial (by applying the same inclusion/exclusion criteria as in the studies), was created from the SEER-Medicare database. The inclusion and exclusion criteria of the RCT, for which it was desired to extrapolate the survival estimates were applied to the SEER-Medicare database to extract a cohort of patients that reflect the RCT target patient population. Metastatic melanoma patients were identified according to the International Classification of Disease for Oncology (ICD-O-3: 8720/3 or 8720/2) and concurrent topology codes (C44.x). Treatment was identified in the Medicare database with line of treatment defined through a stepwise algorithm (Zhao *et al.*, 2014). All patients included in the cohort had continuous enrolment in Medicare (patients enrolled in a health maintenance organization (HMO) or hospice in a certain time window were excluded) to allow for sufficient and continuous follow-up of the study period. See also the supplementary material for further information on the extraction of data from SEER-Medicare, which is provided in the appendix. There were 47 patients on SOC and 84 patients on experimental treatment. At approximately 80 months 46 (98%) of those treated with SOC and 48 (57%) of those treated with experimental treatment had died. In Figure 2, the survival probabilities for the experimental treatment and SOC obtained from the SEER-Medicare database are presented.

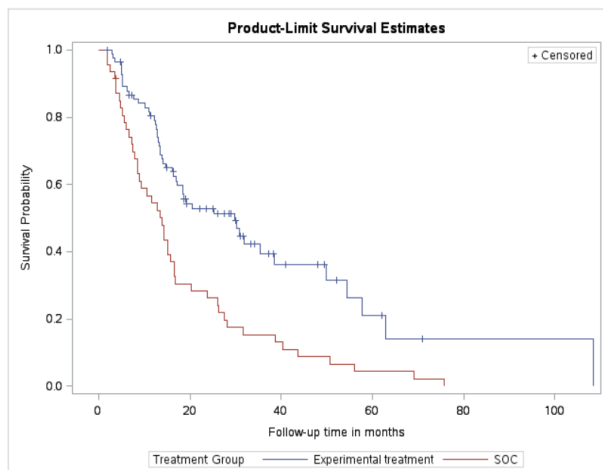

**Figure 2** Survival probability for the experimental and standard of care – SEER-Medicare data

## 2.2 Statistical models

*Parametric models*

As mentioned in the introduction, for purpose of this work, the most commonly used parametric models for the analysis of survival data were used to illustrate the extrapolation strategies. In particular, the Weibull, Gompertz and Exponential as well as the lognormal and log-logistic models were applied to the RCT data. For further details and examples of these parametric survival models we refer to Rodriguez (2005).

Let $T$ be a random variable denoting the survival time then $S(t)=\mathbf{P}\{T>t\}$ is the probability of being alive at time t. In the case of the lognormal model we define $\phi$ as the cumulative standard normal distribution and Log(T) is normally distributed with mean $\mu$ and variance $\sigma^2$ (N($\mu$, $\sigma^2$)). Table 1 presents the functional forms of the parametric models investigated.

**Table 1**  Main parametric survival functions (S(t)) applied and their hazard functions (h(t))

| Model | (S(t)) | h(t) |
|---|---|---|
| Exponential | $\exp(-\lambda t)$ | $\lambda$ |
| Weibull | $\exp(-(\lambda t)^p)$ | $\lambda p(\lambda t)^{p-1}$ |
| Lognormal | $1-\phi[(\log(t)-\mu)/\sigma]$ | $((Exp[-1/2((\log(t)-\mu)/\sigma)])/t(2\pi)^{1/2}\sigma))/S(t)$ |
| Log-logistic | $1/(1+(\lambda t)^p)$ | $pt^{p-1}\lambda^p/(1+(\lambda t)^p)$ |
| Gompertz | $\exp(\lambda(1-\exp(pt)))$ | $\lambda\exp(pt)$ |

The parameters $\lambda$ (scale parameter), $p$ (shape parameter), $\mu$ and $\sigma$ are estimated from the data. In the case of the Weibull function, if $p>1$ this denotes an increasing hazard, while $p<1$ denotes decreasing hazard. When $p=1$ the Weibull model reduces to an exponential model, and as such it can be considered a more general version of the exponential model. The Gompertz model assumes that the survival decreases over time at an exponential rate. As is the case with the Weibull function, $p<1$ and $p>1$ correspond with decreasing and increasing hazard and when $p=1$ the model reduces to an exponential model.

In situations where the distribution of the survival times is skewed and the variability is uncertain, or high, the lognormal distribution may be an appropriate model to use. For such a model the logarithm of the residual errors is assumed to come from a normal distribution.

It is assumed that the assumptions underlying a survival analysis are inherited from the RCT data and still hold due to the independently selected additional data (SEER-Medicare data and long term summary measures reported in literature). In particular, it is assumed that the proportional hazards assumption is met, i.e. the ratio of the hazards between two individuals is constant over time, the observations are independent (by design) and the linear relationship between the covariates and survival is linear.

Sponsors do not routinely use nonparametric models for extrapolation of RCT data as was seen in Latimer (2013). As a result, a nonparametric model was also used to illustrate how such a model could be used. In contrast to parametric modelling a nonparametric model does not assume a specific distribution for the data. In particular, a local linear model was chosen as it is very flexible and can accommodate any number of empirical distributions, which may be of particular interest in the tail of the RCT where there is great uncertainty due to the presence of censored data. The following section describes the nonparametric model in more detail.

*Nonparametric model*

In addition to the applied parametric models, we applied a so called non-parametric model based on local linear regression. This model is referred to as nonparametric because it does not *a priori* assume a shape of the response curve (Li and Racine, 2004). This model was selected because it is relatively easy to understand and is based on a basic mathematical principle that any function can be estimated by a series of local linear functions, e.g. Newton's method for estimate the extreme values on a curve (Clapham and Nicholson, 2014).

In summary, assuming we need to estimate an unknown function f(t) in the interval [a,b]. The interval [a,b] can be split in finite intervals starting from $[a,x_i]....[x_n,b]$ of chosen bandwidth. The bandwidth of each interval can be arbitrarily chosen. However, the accuracy of the estimation will depend on the size of the chosen bandwidth, i.e. the smaller the bandwidth the more accurate the prediction but more accurate prediction requires more iterations. If the bandwidth is chosen such that the length is equal to the distance between two subsequent data points then the model will connect the data points. The estimate of the function in the time interval [a,xi], $\bar{f}_{(t_{[a,x_i]})}$, is a linear model with slope $[s(a)-s(x_i)/|a-x_i|]$ for a decreasing function. Here s(x) denotes the response value at (time) x.

In general, assuming that *S(t)* is the survival time at time *t*, the model as proposed by Li and Racine (2004) is as follows:

$$S(t) = g(x(t))$$

where *g(x(t))* is a local linear fit applied to a range of data points in time and estimated directly from the data observed and *x(t)* is the vector of regressors, e.g.time.

Models were fitted in R (R core team, 2014) and WinBUGS (Spiegelhalter *et al.*, 2003). Goodness of fit for the parametric models was evaluated through assessment of the log likelihood test statistic or the Deviance Information Criterion (DIC) in WinBUGS. The log likelihood for the nonparametric model was calculated using the conditional density. A lower value indicates a better fit. The models were additionally assessed through visual inspection.

The following sections outline how these models were applied to the example data.

### 2.3 Non-informative extrapolation

The parametric statistical models described above and the non-parametric model were fit to available RCT data by regression analyses. The regression model was extended from 48 months (the length of follow-up in the RCT) to 80 months, assuming no data were available from 48 months to 80 months.

The case when there are no long-term follow up data available, will be referred to as non-informative extrapolation.

### 2.4 Extrapolation using re-created IPD from published KM curve and long-term follow-up study as reference

Individual patient data and OS probabilities were recreated from published KM curves (Altomonte, 2013) using the DigitizeIt software (http://www.digitizeit.de/) and R (R core team, 2014) code from Guyot *et al.* (2012). Guyot *et al.* (2012) described an algorithm for digitizing published KM curves back to KM data by inverting the KM equations, using the number of events and numbers at risk (if available). They compared their reproduced survival probabilities, including median survival times and hazard ratios with the published results. It was shown that the reproduced results were comparable to the original results. However, accuracy of the reproduced results was reasonable only if at least the number of patients at risk or the total number of events were reported in addition to the KM curves. In Altomonte (2013) the KM curves as well as the number of patients at risk were reported. As a result, we were able to apply the methodology described by Guyot *et al.* (2012) to the Altomonte published results and generate survival probabilities from the Altomonte published results to use in our attempt to extrapolate the results from the RCT.

Long-term survival results published from a large registry study (Joosse *et al.*, 2011) was used to assess the plausibility of the extrapolation. This was done through visual inspection. Visual inspection was also performed in the approaches described in Sections 2.5, 2.7 and 2.8. For the constrained regression analyses (Section 2.6) the long term survival was used to constrain the applied models.

In the remainder of this document the long-term survival results from the registry study (Joosse *et al.*, 2011) is referred to as "LT follow-up study". Only the best fit parametric models from the non-informative extrapolation, i.e. log-logistic and lognormal models and the non-parametric model are presented to illustrate this and the following approaches.

### 2.5 Extrapolation using SEER-Medicare data and LT follow-up study as reference

Data from the SEER-Medicare database were used to extrapolate the results from the RCT, based on the selected parametric and non-parametric survival models. IPD were generated from the life-table estimates (which included the survival probability, time (months), number of patients at risk and number of patients censored) obtained from the SEER-Medicare patient cohorts using a minor modification of the code provided by Guyot *et al* (2012). The results from the period to be extrapolated were merged (blended) with the RCT data by adding the SEER-Medicare data to the RCT data to create one dataset. This dataset was then used for extrapolation using the models described in Section 2.2. The available follow-up time from the SEER-Medicare data was less than 80 months for SOC and about 110 months for experimental treatment. Using this follow-up time in the extrapolation range may reduce the uncertainty around the estimates obtained from the extrapolation. As a result, the extrapolations have been carried out with and without the SEER-Medicare data to investigate the impact of including these data in the extrapolation strategy.

Because of the uncertainty of the RCT data at the end of the follow-up period due to censoring, the last 10%, 20% or 30% of the RCT data at the end of the follow-up period was removed. The analyses presented here excluded 10% of the RCT data at the end of the follow-up period. Removing more data resulted in a poor statistical fit and is not recommended (Zwiener *et al.,* 2011).

The most commonly reported period for assessment of long-term survival (Queiroga *et al.*, 2003; Kelly and Halabi, 2010) is around 5 to 6 years (72 months). To provide an estimate of long-term overall survival the area under the survival curve restricted to 72 months of follow-up was estimated, i.e. $AUC_{0-72}$, together with its standard error. A restricted AUC measure is often used in Health Technology Assessment (HTA) groups to establish the long-term survival associated with a treatment (Tappenden *et al.*, 2006).

Note that for illustrative purposes, and due to the availability of data, the approach using the SEER-Medicare data is demonstrated with the control arm of the RCT only. However, the methodological principles apply to any treatment arm in an RCT, subject to the comparability of patients in both the RCT and RWD cohort.

### 2.6 Extrapolation using SEER-Medicare data and Bayesian power model constrained to the long-term follow-up study

Let $\alpha_0$ be the weight of the historical data relative to the current study ($0 \leq \alpha_0 \leq 1$), and L($\theta$|D) the general likelihood function for a regression model with parameter vector $\theta$, given data D of the current study. Let $\pi_0(\theta)$ be the prior distribution for theta before the historical data Do is observed. Then the power prior distribution of $\theta$ (for the current study) is defined as (Chen and Ibrahim, 2006):

$$\pi(\theta \,|\, D_0, \alpha_0) \propto L(\theta \,|\, D_0)^{\alpha_0} \pi_0(\theta)$$

Chen and Ibrahim (2006) demonstrated the advantage of a power prior approach especially when multiple historical datasets are available to estimate the power parameter. Ibrahim *et al.* (2015) describe the advantages of using power prior approach over other informative priors in general linear models, survival models and random effect models. The advantages described include:

(i) The techniques to show adequacy of $\pi(\theta|D_0, \alpha_0)$ are the same as those for showing adequacy of the posterior distribution with likelihood function L($\theta$|D) and prior $\pi_0(\theta)$, using likelihood theory

(ii) Variable selection and general model selection problems can be investigated using varying (power) prior distributions

(iii) The asymptotic properties resulting from the likelihood theory also hold for the power prior, since the power prior is a likelihood raised to a power ($\alpha_0$)

The rest of this section describes how the Bayesian power prior model was constrained using the long-term follow-up study information in order to estimate long-term survival.

Let S($t_i$) be the survival time at time $t_i$ and S($t_{obs}$) is the observed survival time at time $t_{obs}$ from the long-term follow-up study. The median survival time observed from the long-term follow-up study was

80 months. Let also, $\sigma_{obs}^2$ denote the observed variance from the long-term follow-up study. Then the lognormal constraint model (Guyot *et al*., 2014) is:

$$S(t_i) = 1 - \phi[(\log(t_i) - \mu)/\sigma], i = 1,..,72$$

Subject to the constraint

$$S(t_{obs}) = S_{obs}$$

Similarly, the Log-logistic constraint model is:

$$S(t_i) = 1/(1 + (\mu t_i)^\tau), i = 1,..,72$$

and ensuring the survival function is constrained to the observed long-term time and survival

$$S(t_{obs}) = S_{obs}$$

For varying levels of α, the precision of the estimates of survivals were defined as follows (assuming τ=1/σ for the log-logistic model):

$$S^*[t_j] = \left(1/\sigma_{obs}^2\right)^\alpha$$

In practice, the choice of the level of α will depend on the confidence and acceptability of the (precision of) real world summary data. For this example, we allowed α to range from 0.001 (low precision) to 1. As a sensitivity analysis α was also set at 2 (higher precision). The scenario where α =1 coincides with the observed variability from the published long-term registry study, i.e. accepting it at "face value".

The prior distributions for μ and τ were assumed to be a normal (N(0,0.01)) and a gamma distribution (Γ(0.001,0.001)) respectively.

The Bayesian power model was applied with (SOC only) and without the SEER-Medicare registry data to illustrate the impact of the RWD and the applied methodology. The long term lymph node metastasis survival result by Joosse *et al*. (2011) was used to constrain the regression models (see also Sections 2.1 and 2.4).

### 2.7    Extrapolation using SEER-Medicare data and BMA

An advantage of the Bayesian Model Averaging approach is that it accounts for model uncertainty (Hoeting *et al*., 1999). For the two best fitting survival models the lognormal and logistic, a BMA approach (Jackson *et al*., 2009) was applied. Let $S_{LN}(t)$ denotes Lognormal survival function and let $S_{LL}(t)$ denote the log-logistic survival function. Determine weights $W_1$ and $W_2$ such that the overall survival function is the weighted average of the individual components:

$$S(t) = W_1 S_{LN}(t) + W_2 S_{LL}(t)$$

With $W_i$ being the weighted average of the Deviance Information Criterion (DIC) obtained from each individual ($S_{LN}$ or $S_{LL}$) model fit.

This approach can be extended to incorporate other plausible survival models as appropriate.

### 2.8 Sensitivity analyses of chosen weights in the BMA approach

To investigate the acceptability of the weights, sensitivity analyses were performed using different weight options obtained from fitting the models in WinBUGS. In the first option, the Weights were estimated as the weighted average of the restricted Area under the curve up to 72 months, $AUC_{0-72}$, obtained from the individual model fit.

A weighted average of the estimated location parameter ($\mu$) from each survival function were also used as weights. Based on the asymptotic properties of the log-logistic and lognormal distributions (Dey and Kundu, 2010) such an approach appears to be valid for large sample sizes. Dey and Kundu also provided correction factors for estimating the probability for correct selection of each model in cases when the sample size is smaller. These correction factors were not further investigated in the context of this research

As described in Section 2.2 and 2.5, $AUC_{0-72}$, together with its standard error was used to describe long term survival. A model with a lower Deviance Information Criteria (DIC) as estimated by WinBUGS (Spiegelhalter *et al.*, 2003) was considered a better model.

## 3 Results

### 3.1 Non-informative extrapolation

The 5 parametric models seem to fit the data reasonably well. The log-likelihood test statistic was lowest for the lognormal and log-logistic models indicating a better fit. Table 2 presents the results of the log-likelihood test statistic for the selected models.

**Table 2** Model fit results for selected models applied to the RCT data.

| Model | -2xloglikelihood (-2LL) |
|-------|--------------------------|
| Exponential | 3120.7 |
| Weibull | 3119.8 |
| Lognormal | 3068.0 |
| Log-logistic | 3067.2 |
| nonparametric | 2100.8 |

The results of the extrapolation for the lognormal and log-logistic models in the absence of long term follow-up data are presented in Figure 3a. The plausibility of the extrapolated survival curve beyond the RCT data is uncertain, especially when benchmarked against the results of the long-term registry study (Joosse *et al.*, 2011). The estimate (SE) for the restricted AUC from 0 to 72 months for SOC was 15.7 (1.2) for the best fitting log-logistic regression model.

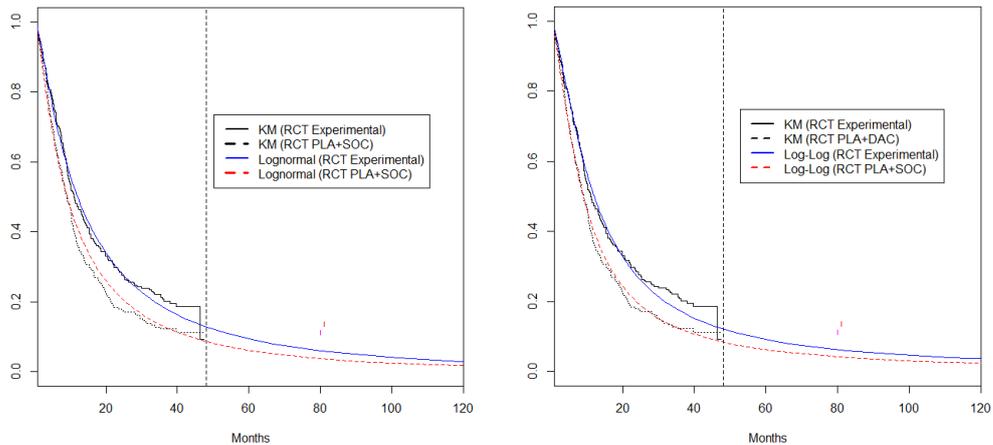

**Figure 3a** Left panel: Non-informative lognormal extrapolation of RCT data. Right panel: Non-informative log-logistic extrapolation of RCT data

With regards to the nonparametric regression model, it was observed that this model fits the RCT data extremely well as confirmed by the smallest log-likelihood of all the models. However, due to the lack of follow-up data and the number of censored observations in the tail of the observed RCT data, the extrapolation using this model when there is no follow up data is inaccurate. If follow-up data and external data are available in a reasonable number of time intervals the nonparametric model would be a good alternative to use for extrapolation. However, if data are sparse the nonparametric model may not be appropriate. As a result, this model was not taken further in the subsequent evaluations but was presented here for illustrative purposes. Figure 3b is presented as an illustrative example of what can be expected if this model was to be used in a constraint extrapolation with sparse data. The long term study (Joosse et al. 2013) is the only available data for the period to be extrapolated and the survival probability is at a much later time point (80 months) than the follow-up period of the RCT (48 months). As a result, the linearity as imposed by the nonparametric extrapolation in the extrapolated interval, is questionable.

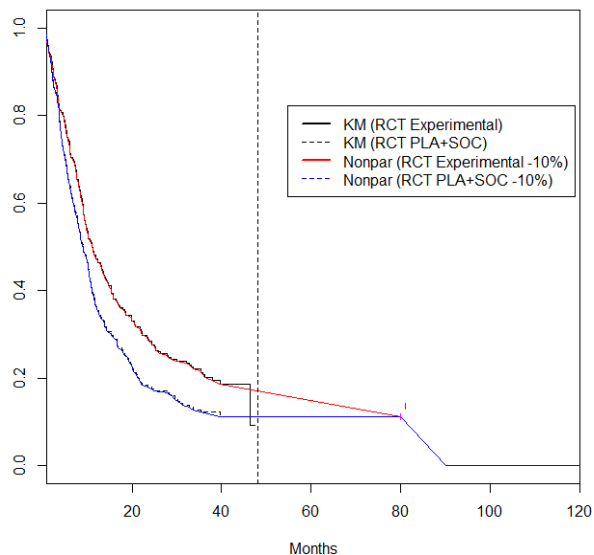

**Figure 3b** Nonparametric extrapolation of RCT study data including published RWD from long-term study (Joosse et al. (2011))

Non-informative extrapolation may be useful when there is real world data available in a reasonable interval in the extrapolation range and the model fit the RCT data reasonably well. However, extrapolation is unreliable in the absence of long-term data. Due to censoring, the follow-up data of the RCT are also uncertain which may have an impact on the quality of the extrapolation. In line with the recommendations from Latimer (2013), 10% of the follow-up data was removed prior to extrapolation in the remainder of the paper.

### 3.2 Extrapolation using re-created IPD from published KM curve and long-term follow-up study as reference

The recreated IPD from published KM-curve and the resulting KM curve are presented in Figure 4.

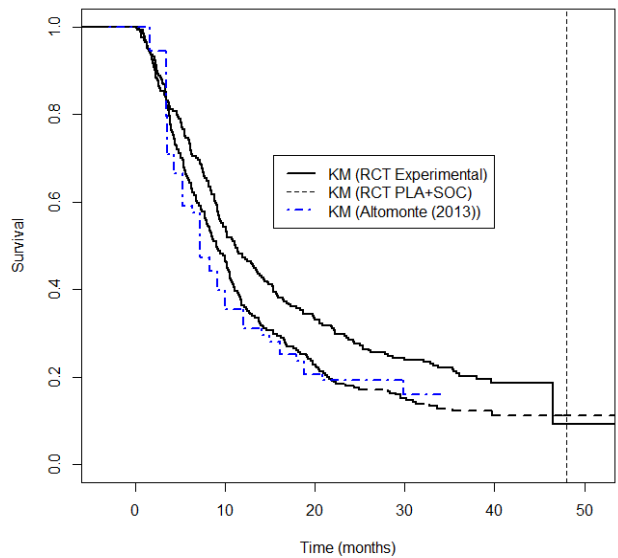

**Figure 4** Digitised Kaplan-Meyer curve of real-world study (Altomonte (2013))

The published results (Altomonte, 2013) approximated those from the patients receiving SOC. However, the follow-up time reported in the published study was similar (even shorter) to the RCT follow-up (48 months). Therefore, the re-created IPD from the KM curve would not be useful for extrapolation purposes and, in this case, digitizing the published Kaplan-Meier curve did not result in useful data for extrapolation. This approach is presented here as an option to obtain external data for extrapolation and could be useful in case the follow-up time of the published KM estimates are longer than the follow-up time of the RCT.

### 3.3 Extrapolation using SEER-Medicare data and long-term follow-up study as reference

Blending of RWD (SEER-Medicare data) with RCT data (by adding the estimated RWD survival probabilities from the extrapolation interval to the survival probability data from the RCT) resulted in a reduction in the uncertainty of the long-term survival estimate ($AUC_{0-72}$) compared to the non-informative extrapolation. The estimate (SE) was 14.8 (1.1). However, this reduction in the standard error should be interpreted with caution and should be considered in conjunction with the appropriateness of the RWD. Based on the long-term summary results (Joosse *et al.* 2011) and the data obtained from the SEER-Medicare registry, clear judgement is needed regarding the acceptability of RWD. For example, the SEER-Medicare registry provided long-term data useful for extrapolation of the SOC arm. However, the early follow-up period (up to 48 months) would appear to differ from the RCT data as is illustrated in Figure 5. Moreover, the extrapolation results of the SOC arm does not approximate the results from Joosse *et al*. (2011), which showed superior survival probabilities at 80 months, compared to the probable time course of the RCT extrapolation.

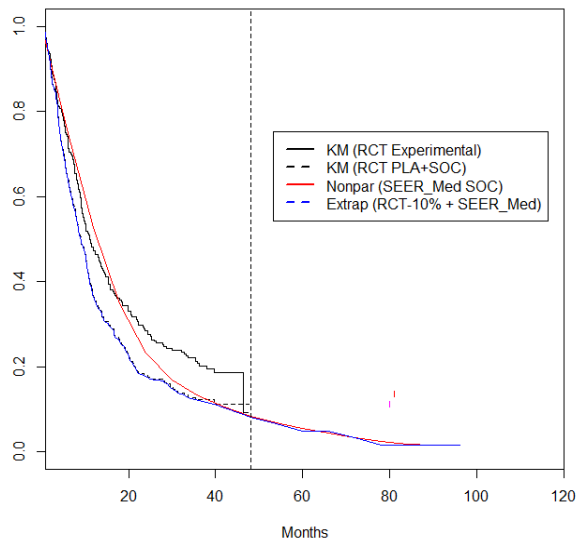

**Figure 5** Extrapolation of RCT data using SEER-Medicare data

### 3.4 Extrapolation with and without SEER-Medicare data and Bayesian power model with constraint on the long-term follow-up study

*Without SEER-Medicare:*

The $AUC_{0-72}$ (SE) using the lognormal model to model the survival of patients on SOC arm was 17.8 (1.1), 17.0 (1.0) and 17.0 (0.9) for α=0.2, α=1.5 and α=2, respectively. This illustrates a further reduction in the uncertainty in the long-term survival estimate compared to the previous approach. Figure 6 illustrates the model fit for varying levels of alpha (left graph) and for the best fitting levels of α (right graph) for both treatment arms. Constraining the model to the summary RWD (Joosse et al, 2011) results in the extrapolation curve departing slightly from the RCT portion of the data, as the extrapolation model aims to approximate the summary RWD. This could be indicative of what was observed previously, that the 80 month survival probabilities of the published RWD (Joosse et al. 2011) are superior to the possible long term survival probabilities of the SOC arm.

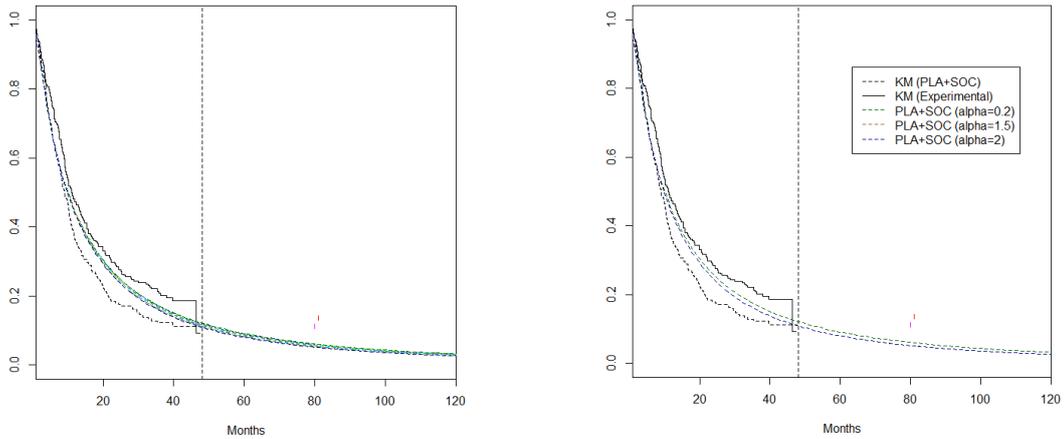

**Figure 6** Extrapolation of SOC with the Bayesian power prior approach without SEER-Medicare data for all levels of the power component (left panel) and selected levels of the power component (right panel)

For the patients receiving experimental treatment, the $AUC_{0-72}$ (SE) for $\alpha=0.001$, $\alpha=0.5$, $\alpha=0.6$, $\alpha=1/2$ was 20.9 (1.2), 20.8 (1.2), 20.7 (1.1) and 20.3 (1.1), respectively. Figure 7 illustrates the results of the Bayesian power prior approach with constraint for the experimental treatment arm and illustrates how well the methodology worked when the long-term registry results (Summary RWD from the long-term registry study (Joosse *et al.*, 2011)) is an achievable target to extrapolate towards.

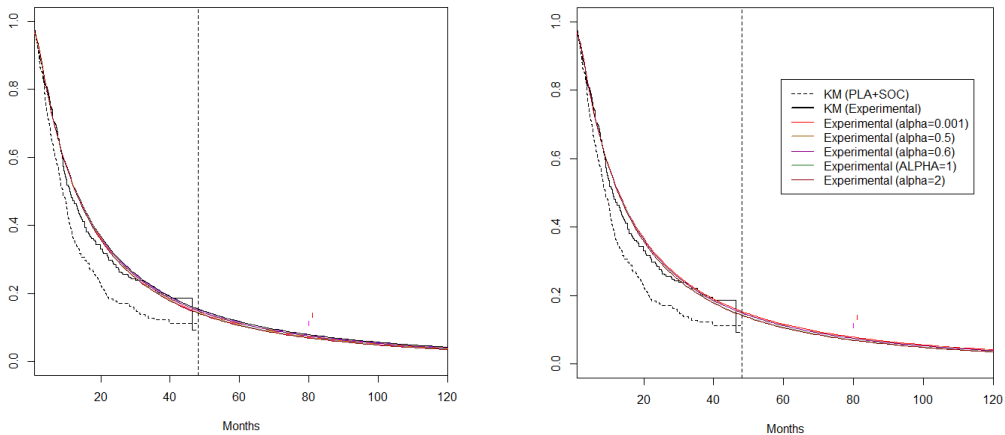

**Figure 7** Extrapolation of Experimental treatment with the Bayesian power prior approach without SEER-Medicare data for all levels of the power component (left panel) and selected levels of the power component (right panel)

*With SEER-Medicare (SOC only):*

The $AUC_{0-72}$ (SE) were 18.1 (1.2), 17.3 (1.0) and 17.3 (1.0) for $\alpha=0.2$, $\alpha=1.5$ and $\alpha=2$, respectively. The results indicate that as the precision in the RWD (Joosse *et al.*, 2011) decreases (i.e. an increase in the variability of the long-term RWD estimate) the RCT data are better fit with a marginal worsening of the uncertainty in the long-term survival estimate compared to when the registry data was not used. However, using the registry data resulted in lower variability in $AUC_{0-72}$ compared to the non-informative extrapolation case.

The results were similar for the log-logistic case, with the results for the log-logistic case being marginally better than those for the lognormal case.

For this approach, the reliability of the results of the extrapolation is dependable on the acceptability of the estimate and variation of the available long-term RWD (Joosse *et al.*, 2011). Whether or not the OS result of the RCT when extrapolated can realistically approximate the long-term estimate is also a point to consider.

### 3.5    Extrapolation using SEER-Medicare data and BMA

The Bayesian model averaging approach reduced the uncertainty of the long-term survival estimate the most compared to all the previous methods, with or without the use of the SEER-Medicare registry data. Without the use of the registry, the $AUC_{0-72}$ (SE) for SOC was 15.1 (0.76) and for patients on experimental treatment the $AUC_{0-72}$ (SE) was 20.9 (0.97). The log-likelihood statistic (-2LL) was 3306.7 and 1627.8 for the lognormal and log-logistic model respectively. These estimates were also lower than those from previous models, indicating improved fit with reduced uncertainty.
When the registry data are blended (see Section 2.5) with the RCT data (SOC only), the $AUC_{0-72}$ was 15.8 with a marginally larger increase in uncertainty (0.78). This estimate is still lower than the previous methods, indicating superiority of the BMA approach compared to the previous methods, at least in terms of uncertainty. The likelihood ratio test statistic was 3341.6 and 1646.4 for the lognormal and log-logistic model respectively. These were marginally larger than without the SEER-Medicare data but still lower than those observed before. However, contrary to what was observed in the power prior approach, the BMA approach fit the RCT data reasonably well and the extrapolated results seem to approximate the results published in Joosse et al. (2011), especially for the experimental arm.

Table 3 provides a summary of the results of $AUC_{0-72}$ for SOC with and without registry data for the lognormal model.

**Table 3**    Summary of results of $AUC_{0-72}$ (SE) for SOC with and without registry data for the lognormal model.

| Method | $AUC_{0-72}$ (SE) |
|---|---|
| Non-informative extrapolation | 15.7 (1.2) |
| Extrapolation using registry data and long-term follow-up study | 14.8 (1.1) |
| Bayesian power model without SEER-Medicare ($\alpha$=2) | 17.0 (0.9) |
| Bayesian power model with SEER-Medicare ($\alpha$=2) | 17.3 (1.0) |
| Bayesian model averaging without SEER-Medicare | 15.1 (0.76) |
| Bayesian model averaging with SEER-Medicare | 15.8 (0.78) |

Future work in this area could include a combined BMA and Bayesian power prior approach to investigate whether combining these approaches will improve the extrapolation results compared to the BMA and Bayesian power prior approach individually.

### 3.6    Sensitivity analyses of chosen weights in the BMA approach

When the weighted average of the restricted Area under the curve up to 72 months were used as the weights in the BMA approach, the estimate for long-term survival was 16.8 (0.86) for SOC and 21.7 (1.01) for patients receiving experimental treatment. In the case that $\mu$ of the individual model fits were used as weights, the $AUC_{0-72}$ (SE) was 16.7 (0.88) for SOC and 21.9 (1.05) for patients receiving experimental treatment.
The results presented here are for the case including the SEER-Medicare registry. These results confirm that the BMA approach was able to reduce the uncertainty of the extrapolation of overall survival compared to the other approaches presented.

## 4    Discussion

We considered different approaches that could be used to extrapolate results of RCTs to estimate long-term survival, with or without external data. In the absence of long-term follow-up data, non-informative extrapolation may be considered. However, due to loss to follow-up, the fit in the tail of the

RCT follow-up may be inadequate. This was most visible in the non-parametric model, which fit the RCT data particularly well, as confirmed by the smallest log-likelihood statistic, but nevertheless showed a poor fit in the tail of the follow-up period. If a reasonable amount of external data were to be available in the period to be extrapolated then the nonparametric model could be a good alternative to the parametric approaches. A model that can be used for extrapolation should fit both the RCT data as well as the external data collectively. However, as proposed by Tremblay (2015) the first step is to ensure the model fits the RCT data well.

The use of SEER-Medicare registry data allowed for extrapolation of RCT results to establish long-term survival with reduced uncertainty. The overall survival results of the IPD which were obtained by digitizing published KM curves of RWD cohorts (Altomonte, 2013) were consistent with findings from the RCT. However, the follow-up of patients in this data source was similar to those in the RCT so did not allow for long-term extrapolation. By combining the RCT with the SEER-Medicare cohort and using the long-term registry study as a reference the long-term extrapolation was achieved, the uncertainty was reduced and the plausibility of the extrapolation was benchmarked. The magnitude of the reduced uncertainty could be driven by the relatively long follow-up of this example RCT and the uncertainty could potentially be larger when extrapolating trials with shorter follow-up.

As mentioned earlier, for purpose of this work, the most commonly used parametric models for the analysis of survival data were applied to illustrate the extrapolation strategies: the Weibull, Exponential, log-logistic and lognormal. However, other parametric models such as the Gamma, Gompertz, Generalised Gamma or Generalised F could also be considered. The applied nonparametric model was chosen due to its flexibility and computational simplicity. However, other spline models or fractional polynomials could also be applied (Royston & Lambert, 2011).

The BMA approach was superior to the other approaches investigated: it reduced the uncertainty in the long-term estimate of survival with a reasonable fit to the RCT data as indicated by the log-likelihood test statistic. The reduction of uncertainty should be considered in conjunction with the model fit diagnostics and the appropriateness of the external data. In this case study the patients from registry were selected to reflect the RCT population as closely as possible. With respect to the models used, we included only those parametric models that best fit the RCT data. Another approach would be to fit all parametric models considered. However, the models that do not fit the RCT well will have potential to add noise to the BMA approach. The inclusion of all models has greater value for reducing noise when models are adequate but none is better than the other. BMA may reduce the uncertainty from all the individual models. However, Fragoso and Neto (2015) performed a systematic review of the use of BMA in various areas of research and concluded that little consideration has been given as to how to select which models to include in BMA. Further research to evaluate selecting the best models compared with including all models in the BMA approach is warranted to provide guidance or recommendations in the future.

The extrapolation analyses presented here excluded 10% of the RCT data at the end of the follow-up period to remove some of the uncertainty in the tail of the survival distribution due to patients being censored. As part of a sensitivity analysis more data (e.g. 20% or 30% of the RCT data at the end of the follow-up period) were also removed, but this resulted in poor fit to the RCT data, and was therefore not pursued or considered appropriate in this case study. This is consistent with the recommendations by Zwiener *et al.* (2011). A simulation study to evaluate different data cut-offs of the RCTs as well as the registry data could be considered for future research.

A limitation of this analysis was the difficulty to access IPD from multiple registries. When registry data are available only summary statistics are sometimes provided. Although summary statistics are useful, this can limit the analyses that can be performed using RWD, such as the adjustments for patient characteristics that have an impact on their long-term survival, e.g. the Eastern Cooperative Oncology Group (ECOG) status at baseline. Additionally, these extrapolations were based on a cohort of US elderly patients, whose survival may not be generalizable to younger patients. However, this subgroup of patients is considered to represent a key subgroup of metastatic melanoma patients, with the median age of metastatic melanoma diagnosis of 63 years (Howlader *et al.*, 2016). As discussed in the introduction, previous extrapolation studies have been undertaken with SEER data (Guyot et al., 2016). SEER provides data on cancer patients of all ages, but systemic anti-cancer treatment information is limited. Utilizing survival and patient cohorts not segmented by systematic treatment regimens may result in biased results as the extrapolations are based on data from two patient cohorts treated with treatments of different efficacy or mechanisms of action. Moreover, as was described the effects in the SEER-Medicare cohort up to 48 months were different than those observed in the clinical trial. One of the reasons for this could be poor adherence to medication or different treatment strategies, as these have been reported to be the main reasons for differences in effects in RCTs compared to real world data (Carls *et al.*, 2017; Martin *et al.*, 2005). It is unknown whether patient compliance or different treatment strategies were the reasons for the difference in effects observed in the RCT and SEER-Medicare cohort up to 48 months. This warrants further investigation.

The use of IPD and summary data to extrapolate and to assess the plausibility of the extrapolation was useful in this analysis. The methods that can be applied in practice will depend on which type of data are available, e.g. IPD or aggregate data. However, further research is warranted to evaluate different approaches with multiple RWD cohorts and assess the impact of using summary data. Additional research could include combining Bayesian Model Averaging techniques (Jackson et al. 2009) and constrained regression models (Guyot *et al.* 2014) in combination with the power prior approach on the uncertainty, especially using prior model probability weights based on external (real world) evidence (Abrams & Happich, 2016). In cases where long term survival is established in a disease area, the consistency and reproducibility of the $AUC_{0-72}$ could also be evaluated through simulations as a way to support a claim whether a drug provides evidence of long term survival. The sensitivity analyses confirmed the robustness and adequacy of the Bayesian Model averaging for extrapolation. The asymptotic properties of the lognormal and log-logistic distribution allows the use of the location parameter as a weighting factor. However, using the location parameter as a weighting factor warrants further investigation, particularly in smaller sample sizes.

**Acknowledgements**   The research leading to these results was conducted as part of the GetReal consortium. For further information please refer to www.imi-getreal.eu. This paper only reflects the personal views of the stated authors. The work leading to these results has received support from the Innovative Medicines Initiative Joint Undertaking under grant agreement n° 115546, resources of which are composed of financial contribution from the European Union's Seventh Framework Programme (FP7/2007-2013) and EFPIA companies' in kind contribution.

This study used the linked SEER-Medicare database. The interpretation and reporting of these data are the sole responsibility of the authors. The authors acknowledge the efforts of the National Cancer Institute; the Office of Research, Development and Information, CMS; Information Management Services (IMS), Inc.; and the Surveillance, Epidemiology, and End Results (SEER) Program tumor registries in the creation of the SEER-Medicare database.

The authors also express their gratitude to Rachel Kalf and Amr Makady of the National Health Care Institute of the Netherlands, Zorg Institute Nederland (ZIN) for performing the literature search to identify relevant studies that reported KM curves.

**Conflict of Interest**

*The authors Reynaldo Martina, David Jenkins, Pascale Dequen, Michael Lees and Frank Corvino have declared no conflict of interest.* Keith Abrams has served as a speaker, a consultant and an advisory board member for Amaris, Allergan, Astellas, AstraZeneca, Boehringer Ingelheim, Bristol-Meyers Squibb, Creativ-Ceutical, GSK, Janssen, Merck, Novartis, NovoNordisk, Pfizer, PRMA, and Roche, and has received research funding from Pfizer and Sanofi. *Sylwia Bujkiewicz has served as a speaker and a consultant for Roche. Jessica Davies declares stock ownership and employment with F. Hoffman-La Roche Ltd.*

# Appendix A   Supplemental material relating to the SEER-Medicare cohort

## A.1   Data source

Observational data from the National Cancer Institute's (NCI) SEER-Medicare program was used to extract a cohort of elderly patients with newly diagnosed metastatic melanoma receiving treatment. NCI's program provides the linkage of two large US population-based data sources that provide patient-level information on Medicare beneficiaries with cancer. Briefly, SEER is a United States national cancer registry covering 28% of the population with the aim of providing tracking on cancer incidence and survival. The data provides patient-level information on cancer patients of all ages with data on demographics, tumor characteristics (stage, histology, site location), and patient survival. The registry is linked to the Medicare insurance claims database, which is a nationally funded health insurer in the US, provided to all residents $\geq$65 years of age or diagnosed with end-stage renal disease or other disabilities. 97% of US residents $\geq$65 years of age qualify for Medicare coverage (National Cancer Institute, 2016).

In this study we used a 2015 SEER-Medicare release database, which included SEER diagnosis and death information between January 1, 1995 and December 31, 2012 with Medicare claims through December 31, 2013. Part D (outpatient prescription drug) data are available between Jan 1, 2007 and Dec 31, 2013

### A.2    Cohort selection

In this study we included treated elderly patients ($\geq$66 years of age) diagnosed with metastatic melanoma between January 1, 2004 and December 31, 2011. We extracted patients with an International Classification of Disease for Oncology (ICD-O-3) code 8720/3 or 8720/2 and concurrent topology codes (C44.x) (n=7,149). Stage at diagnosis was extracted from the SEER database, and only patients with metastatic disease at diagnosis were selected. SEER provides stage information using the American Joint Committee on Cancer (AJCC) staging system. Patient selection was cut-off on December 31, 2011 to allow for at least one year follow-up after metastatic diagnosis. Patients with a diagnosis of a primary tumor other than melanoma 6-months prior or any time after melanoma diagnosis were excluded. Additionally, all patients had continuous enrollment in Medicare Part A and Part B (no enrollment in a health maintenance organization or hospice) for at least 1 month prior to diagnosis and during the entire follow-up period to ensure ability to create health history data for each patient. The age cut-off was $\geq$66 years of age at diagnosis to allow for at least 1 year of Medicare data prior to diagnosis.

All patients included in the cohort initiated front-line treatment with the treatment(s) of interest as a monotherapy or combination. This was identified through $\geq$2 Medicare claims after diagnosis date (n=131).

### A.3    Treatment algorithm

Treatment lines were defined based on previous insurance claims studies in metastatic melanoma (Zhao, 2014). First-line treatment was defined as receipt of first systemic anti-cancer treatment consisting of the treatment(s) of interest as monotherapy or combination initiated any time after metastatic diagnosis. At least two claims of the particular agent had to occur within 28 days to be considered. Combination therapies were defined as occurrence of both therapies within the same 28 day window. A new line of therapy was defined as either a 90-day gap in treatment or initiation of a new regimen that does not occur within 28 days of the start of the previous line of therapy.

### A.4    Statistical analysis

Lifetable survival estimates were generated for the patient cohort by treatment regimen to extrapolate the survival estimates from the RCT using the SEER-Medicare data. These survival estimates from SEER-Medicare for a population reflected as closely as possible to the RCT were blended (added, see Section 2.5) with the survival estimates from the RCT for the extrapolations.

**SEER-Medicare Flow Chart**

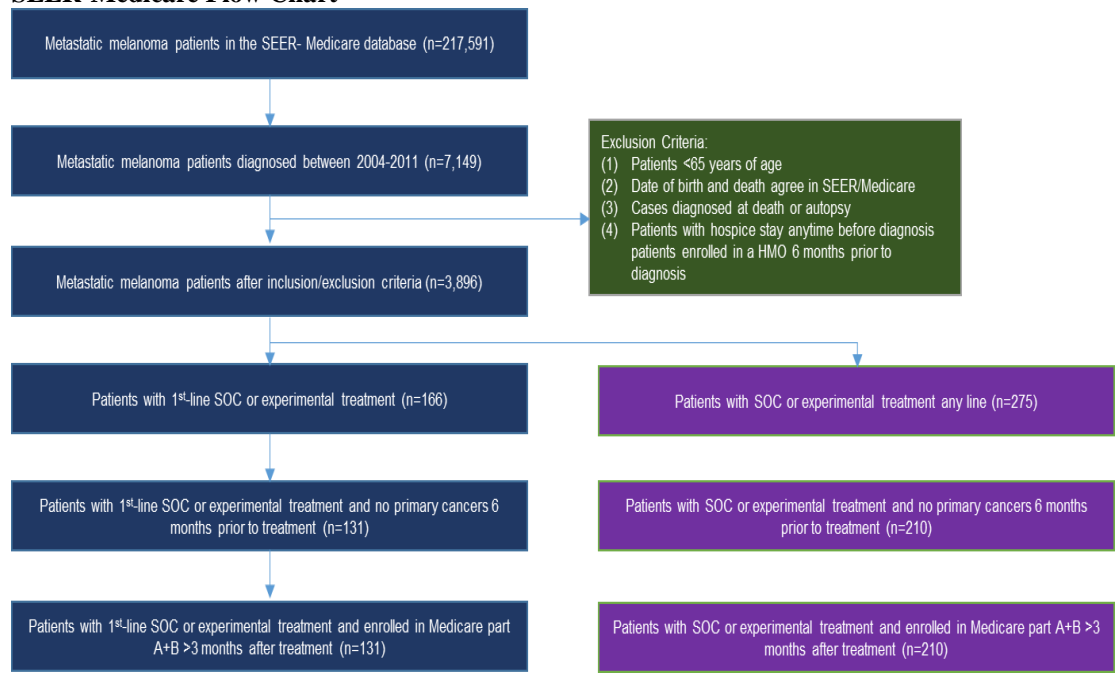